\documentclass[nature,twocolumn]{revtex4}
\usepackage{epsfig,amssymb,amsfonts}

\begin{document}
\title{A proposal for the implementation of quantum gates with photonic-crystal coupled cavity waveguides}
\date{\today}
\author{Dimitris G. \surname{Angelakis}$^{1}$$^*$
}%
\author{Marcelo F. \surname{Santos}$^{2}$}%
\author{Vassilis \surname{Yannopapas}$^{3,4}$}%
\author{Artur \surname{Ekert}$^{1,5}$}%
\email{dimitris.angelakis@qubit.org}
\address{$^{1}$Centre for Quantum Computation, Department
of Applied Mathematics
 and Theoretical Physics, University of Cambridge,
 Wilberforce Road, CB3 0WA, UK}
\address{$^{2}$Dept. de F\'{\i}sica, Universidade Federal de
  Minas Gerais, Belo Horizonte,
30161-970, MG, Brazil}
\address{$^{3}$Condensed Matter Theory Group, Blackett
Laboratory, Imperial College, London, SW7 2BW, UK}
\address{$^{4}$Department of Materials Science, University of Patras,
  Patras 265 04, Greece}
\address{$^{5}$Department of Physics,
             National University of Singapore,
             Singapore 117\,542, Singapore }

\begin{abstract}
Quantum computers require technologies that offer both sufficient
control over coherent quantum phenomena and minimal spurious
interactions with the environment. We argue, that photons confined
to photonic crystals, and in particular to highly efficient
waveguides formed from linear chains of defects doped with atoms or
(quantum dots) can generate strong non-linear interactions which
allow to implement both single and two qubit quantum gates. The
simplicity of the gate switching mechanism, the experimental
feasibility of fabricating two dimensional photonic crystal
structures and integrability of this device with optoelectronics
offers new interesting possibilities for optical quantum information
processing networks.
\end{abstract}

\maketitle
\bibliographystyle{apsrev}

In order to perform a quantum computation one should be able to
identify basic units of quantum information i.e qubits, initialize
them at the input, perform an adequate set of unitary operations and
then read the output~\cite{BEZ}. Here we show that these tasks can
be performed efficiently using photons propagating in the lines of
defects in photonic crystals\cite{Yablo_John,joanno}. These
structures are known as coupled resonators optical waveguides
(CROWs) or coupled cavity waveguides (CCWs) and support efficient
low loss guiding, bending and coupling of light pulses at group
velocities of the order of $10^{-3}$ the speed of
light~\cite{CCW,Poon04}. Qubits can be represented by the ``dual
rail" CROW, i.e. by placing a photon in a superposition of two
preselected lines of defects such that each line represents the
logical basis state, 0 or 1. Quantum logic gates are then
implemented by varying the length and the distance between the CROWs
and by tuning the refractive index in some of the defects using
external electric fields and cavity QED type enhanced non-linear
interactions between the propagating photons\cite{CQED}. We start
with a sketch of the underlying technology followed by a more
detailed description of quantum logic gates and conclude with the
estimation of the relevant experimental parameters.

\begin{figure}
\centering \epsfxsize=6 cm \epsfbox{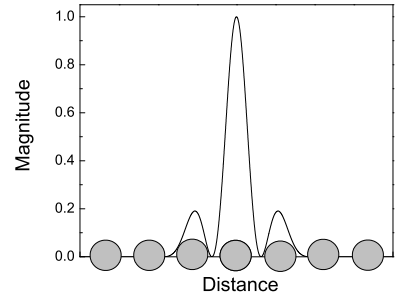} \caption{Snapshot of
a pulse propagating inside a CROW. The field intensity is mostly
localized inside the defects of the
CROW\cite{CCW,Poon04,MZIs,MZIs_exp,FTTAN05}. } \label{pulse}
\end{figure}

Photonic crystals (PCs) are made of ordered inhomogeneous dielectric
media with a dielectric constant spatially periodic on the same
scale as the wavelength of the light propagating in them. They can
exhibit band gaps and also defects like their electronic
counterparts, the semiconductors. Point and line defects can also be
introduced in them with the latter constituting very efficient
waveguides\cite{PBG_fibers}. A point defect introduces a bound state
of the electromagnetic field within the photonic band gap which can
act as a high-Q cavity. Many point defects can be brought together
to form the above mentioned CROWs. A light pulse which enters a
CROW(Fig.~\ref{pulse}),  propagates through a tunnelling/hopping
mechanism between neighboring defects allowing for low dispersion,
small group velocities and efficient
waveguiding\cite{CCW,Poon04,MZIs,MZIs_exp,FTTAN05}.

After preselecting two CROWs and labeling them as 0 and 1 we can
perform an arbitrary unitary operation on the resulting qubit by
concatenating elementary single qubit gates such as the Hadamard
gate and a phase shift gate. The Hadamard gate can be implemented
by bringing the two CROWs of the same qubit closer to each other,
about one lattice constant apart, to allow photons to tunnel
between them. This process, apart from phase factors, is
equivalent to the action of a beam-splitter, or an optical
coupler, in conventional optics. A single qubit phase gate can be
implemented by increasing the length of one of the two CROWs; the
resulting time-delay induces a relative phase shift.

\begin{figure*}
\includegraphics[width=0.5\textwidth]{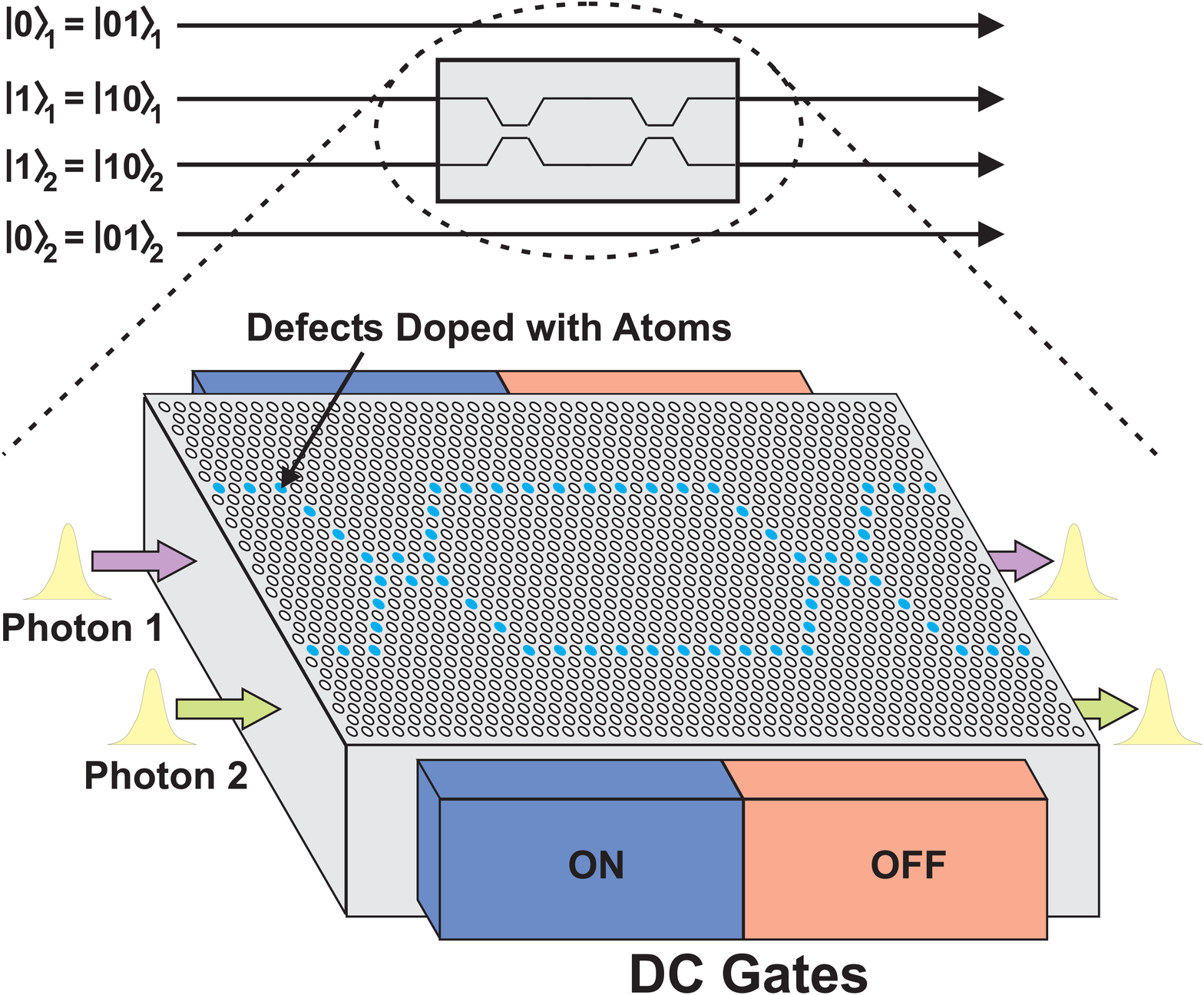}
\caption{The upper part shows a schematic of four coupled cavity
waveguides (CROW) which represent two qubits. The two central
waveguides, belonging to two different qubits, are brought
together in a nonlinear interferometric device which is shown
below the schematic. The device is integrated into a 2D,
micrometer size photonic crystal. The lines of defects, shown in
blue, transfer photons from left to right. The two waveguides are
brought closer to each other right after the entrance and before
the exit of the device, allowing photons to tunnel between them.
The defects in between these two regions are doped with atoms or
quantum dots which can be tuned to be on and off-resonance with
the propagating light by applying an external electric field. An
interplay between the resonant two photon and the dispersive one
photon transitions leads to phase shifts required both for single
qubit phase gates and two qubit controlled-phase gates. The green
and red boxes mark the area with the electric field on and off,
respectively. The field is switched on to induce the nonlinear
phase shift. However, at the end of the quantum gate operation the
field is selectively turned off to the right of the defect where
the phase shift was induced. This is represented by the half green
and half red area, as shown in the picture, and allows photons to
be released back to the propagating modes.} \label{MZI}
\end{figure*}

As an example consider a single qubit interference, i.e. a
sequence: the Hadamard gate, a phase gate, the Hadamard gate. It
can be implemented by a device shown in the lower part of
Fig.(\ref{MZI}), which is a Mach-Zehnder interferometer embedded
in a photonic crystal. The two Hadamard gates correspond to the
two areas in which the CROWs are brought closer to each other.
Relative phase $\phi$ can be introduced by varying the length of
one of the CROWs in the area between the two Hadamard gates. If a
pulse of light is injected into one of the input ports it will
emerge at the one of the two output ports with the probabilities
$\sin^{2}(\phi/2)$ and $\cos^{2}(\phi/2)$, where $\phi$ is the
accumulated phase difference between the two arms. This has been
demonstrated experimentally for 2D CROW structures in the
microwave regime and has been extensively studied for the optical
case~\cite{MZIs,MZIs_exp,Poon04}.

The existing experimental realizations of a photonic crystal MZI had
the phase shift $\phi$ fixed by the architecture, however, one can
also introduce an active phase control. It can be achieved by
placing a medium with tunable refractive index into one of the arms
of the interferometer in between the Hadamard gates. Defects in one
of the arms can be doped with atoms or quantum dots of resonance
frequency $\omega_{ge}$. These two-level systems can be then tuned
to be on and off-resonance with the propagating light of frequency
$\omega$ by applying an external electric field, i.e. by using the
Stark effect. Initially the dopants are far off resonance with the
light pulse, which allows the pulse to enter the CROWs without any
reflections. As soon as the pulse reaches the area in between the
Hadamard gates the electric field is applied bringing the dopants
closer to resonance and inducing a near-resonant dispersive
interaction. When the detuning $\delta=\omega_{ge}-\omega$ is
smaller than both $\omega$ and $\omega_{ge}$ and, at the same time,
much larger than the coupling constant between the dopant and the
light field $\Omega$, i.e. when $\omega_{ge},\omega >>
\delta>>\Omega$, then the combined dopant-light system acquire a
phase proportional to $(\Omega^2/\delta)T$, where $T$ is the
interaction time. Both $\delta$ and $T$ can be controlled and we can
therefore introduce any desired phase shift between the two arms of
the interferometer.

Let us now show how the device shown in Fig.(\ref{MZI}) can be
used to implement a two-qubit conditional phase gate. The two
qubits are represented by four CROWs labelled as $|0 \rangle_{1}$,
$|1 \rangle_{1}$ and $|0 \rangle_{2}$, $|1 \rangle_{2}$
respectively for the first and the second qubit. Only two of the
four CROWs enter the device. They have labels $|1 \rangle_{1}$ and
$|1 \rangle_{2}$ and represent the binary 1 of the first and the
second qubit. Thus the device operates either on vacuum (input $|0
\rangle_{1}|0 \rangle_{2}$), or on a single photon (inputs $|0
\rangle_{1}|1 \rangle_{2}$ and $|1 \rangle_{1}|0 \rangle_{2}$) or
on two photons (input $|1\rangle_{1}|1 \rangle_{2}$). The desired
action of the device, i.e. the conditional phase shift gate, is:
$|0\rangle_1|0 \rangle_2 \rightarrow |0\rangle_1|0 \rangle_2$,
$|0\rangle_1|1 \rangle_2 \rightarrow |0\rangle_1|1 \rangle_2$,
$|1\rangle_1|0 \rangle_2 \rightarrow |1\rangle_1|0 \rangle_2$,
$|1\rangle_1|1 \rangle_2 \rightarrow -|1\rangle_1|1 \rangle_2$.
The device should let the vacuum and one photon states pass
through undisturbed and react only to a two photon state. We can
achieve this by an interplay of dispersive interaction for single
photons and resonant interactions for two photons.

Let us focus only on the CROWs modes that actually enter the
device, i.e. $|1 \rangle_{1}$ and $|1 \rangle_{2}$, and consider
their photon occupation numbers. From now on $|nm\rangle$ means
$n$ photons in mode $|1 \rangle_{1}$ and $m$ photons in mode $|1
\rangle_{2}$. If no phase shift is induced the device affects the
transformation: $|00 \rangle \rightarrow |00 \rangle \rightarrow
|00 \rangle$, $|01 \rangle \rightarrow (|01 \rangle - |10
\rangle)/\sqrt{2} \rightarrow |01 \rangle$, $|10 \rangle
\rightarrow (|01 \rangle + |10 \rangle)/{\sqrt{2}} \rightarrow |10
\rangle$, $|11 \rangle \rightarrow (|20 \rangle - |02
\rangle)/{\sqrt{2}} \rightarrow |11 \rangle$, where the first and
the second arrow correspond to the action of the first and the
second Hadamard gate, respectively. All we need is a nonlinear
medium in between the Hadamard gates such that the states $|00
\rangle$, $|01 \rangle$ and $|10 \rangle$ do not change, while the
states $|20 \rangle$ and $|02 \rangle$ both acquire the same phase
$\pi$.

\begin{figure}
\centering \epsfxsize=3 cm \epsfbox{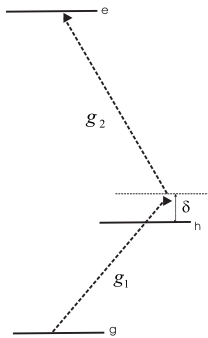} \caption{The
relevant energy levels of the dopants. A photon of frequency
$\omega$ is equally detuned from $\omega_{gh}$ and $\omega_{he}$
($\pm\delta$) and undergoes a dispersive interaction with the
dopants. However, a two photon pulse is resonant with the energy
separation between the levels $g$ and $e$, i.e. $2\omega =
\omega_{gh}+\omega_{he}$, and undergos the resonant interaction.}
\label{atom}
\end{figure}

Following our scheme for the tunable single qubit phase gate let us
now consider dopants with three-level configuration, with electronic
levels $g$, $h$ and $e$ forming a cascade with transition
frequencies $\omega_{gh}$ and $\omega_{he}$. The two transitions
couple linearly to the hopping photons through electro-dipole
interactions, as shown in Fig.(\ref{atom}). We place the dopants in
both arms of the interferometer. A photon of frequency $\omega$ is
symmetrically detuned from $\omega_{gh}$ and $\omega_{he}$ so that
$\delta=|\omega_{gh}-\omega|=|\omega_{he}-\omega|>>g_1,g_2$, where
$g_{1},g_{2}$ are the corresponding coupling constants for the two
transitions. Thus a single photon can only undergo a dispersive
interaction with the dopants. However, a pulse with two photons is
resonant with the energy separation between the levels $g$ and $e$,
i.e. $2\omega = \omega_{gh}+\omega_{he}$, and undergos the resonant
interaction. This can be quantified by the effective Hamiltonian,
extensively studied in the theory of micromasers~\cite{BRH87},
\begin{equation}
H_{\mbox{eff}}=\frac{g_1^2}{\delta}\sigma_{gg} (a^\dagger a) +
\frac{g_2^2}{\delta}(\sigma_{ee} a a^\dagger) + \frac{g_1
g_2}{\delta}(\sigma_{ge} {a^\dagger} ^2 + \sigma_{eg} a^2),
\label{eq:eff}
\end{equation}
where $a^\dagger$, $a$ are the photon creation and annihilation
operators and $\sigma_{ij}= | i\rangle\langle j|$ with $i,j=g,h,e$
are the corresponding atomic operators. The first two terms
describe the dispersive interaction and the third term the
two-photon resonant interaction.

If the dopant is initially in level $|g \rangle$  then the joint
dopant-field state evolves, after time $t$, to~\cite{BRH87}
\begin{eqnarray}
&&|g \rangle |00 \rangle \rightarrow |g\rangle |00 \rangle, \\
&&|g \rangle |01 \rangle \rightarrow e^{-i \varphi}|g \rangle |01 \rangle, \\
&&|g \rangle |10 \rangle \rightarrow e^{-i \varphi}|g \rangle |10 \rangle,\\
&&|g \rangle |20 \rangle \rightarrow e^{2 i \varphi} [\cos{\kappa
t}|g \rangle |20 \rangle +
\sin{\kappa t} |e \rangle |00 \rangle],\\
&&|g \rangle |02 \rangle \rightarrow e^{2 i \varphi} [\cos{\kappa
t}|g \rangle |02 \rangle + \sin{\kappa t} |e \rangle |00 \rangle],
\end{eqnarray}
where $\kappa=\frac{(g_1 g_2 \sqrt{2})}{\delta}$, and $\varphi =
\frac{(g_1)^2 t}{\delta}$. For $\kappa t = \pi$, the two-photon
interaction completes a full Rabi oscillation, acquiring a total
phase $\phi = \pi + 2 \varphi$, where $\varphi = \frac{g_1
\pi}{g_2 \sqrt{2}}$. The ratio $g_1/g_2 = 2 \sqrt{2}$ gives
$\varphi = 2 \pi$ which means that the two-photon state acquires a
minus sign while the remaining states are brought back to their
originals. Under these conditions, the time-evolution showed above
reproduces an instance of a two qubit conditional phase shift
gate.

For the photonic quantum computation, as described above, to be
experimentally feasible we need doped 2D crystal structures of high
quality, strong dopant-photon coupling, and reliable single photon
sources together with efficient photo-detectors. We expect these
requirements to be available with current near future technology.
More specifically:

2D crystals with leaking losses from the CROW as small as a few
percent and defect quality factors of the order of $10^{5}-10^{6}$
(in the near optical frequency regime) should be fabricated soon as
indicated by both the experimental progress in 1D and detailed
theoretical simulation in 2D ~\cite{CCW,MZIs, MZIs_exp}. Doping
active elements in defects in the form of quantum dots  and strong
coupling regime has also been implemented for single defects.
Although technologically challenging, the extension to many coupled
defects should be feasible soon\cite{Strongcoupling, FTTAN05}. In
our scheme for a quality factor of $10^6$ we get a typical
time-scale for undisturbed coherent quantum operations to be of the
order of $T_{1}=1ns$. Both the phase shift operation and the two
photon nonlinear phase shift can be performed within a time period
which is shorter at least by one order of magnitude. The coupling
constant $g$ for the individual atom-photon coupling, for example
for the D2 atomic transition ($852$~nm) of a doped atom of
$^{133}$Cs, is of the order $3\times
10^9$~Hz\cite{CQED,Strongcoupling}. The maximum induced phase is
$\sqrt Ng^2T_{1}/\Delta $ where $N$ is the number of dopants in the
defects. If $\Delta\approx 3\times 10^{10}$~Hz and $N\approx 100$
dopants, then the time required to induce any phase between $0$ and
$\pi$, is roughly $0.1$~ns. Similarly for the two photon nonlinear
phase shift; the two photon Rabi frequency is proportional to $\sqrt
N g_1g_2/\Delta$ and $g_{1}$ is very close to $g_{2}$, and both are
of the order of $3\times 10^{9}$Hz. With the same typical value of
$\Delta$ we get the gate operation time to be of the order of
$0.1$~ns. We note that these figures can be improved by adding more
dopants to the defects making the coupling stronger or by
fabricating higher quality factor defect cavities. Note that for the
case of quantum dots, dipole moments are larger  and  they will thus
couple stronger to the field. However tuning between  dots in
different defects might be a problem in that case. Lastly, the
switching time of the external gates depends on the photon crossing
time, which for a typical CROW group velocity of the order of
$10^{-4}$c is of the order of nanoseconds. Thus the required
switching of the external electric fields should be performed on a
timescale from nanoseconds to tens of picoseconds which is within
current technology. We would also like to add here that the shifting
could also occur by applying a slightly detuned laser field (AC
stark shift), coupled to some other atomic level far from the
hopping photon resonance and the atomic states into consideration.
The size of the accessible device area will then be reduced to the
focus area of the pulse which could be of the order of the
wavelength, i.e. a few microns.


Decoherence due to coupling of the atoms to the vibrations of the
medium is expected to negligible for the case of a suspended atom
(or cold atomic cloud) inside or close to the surface of the defect.
This could be achieved through the lowering of a trap for example on
top of the defect.

Another source of errors could be from the presence of disorder of a
CROW in the optical/infrared regime. Current results in 1D and 2D
show that is possible to implement CROWs in the near
optical/infrared with very low disorder, in good agreement with
theory\cite{CCW,MZIs,MZIs_exp,Poon04,FTTAN05}. Intense efforts to
implement coupled CROWs in 2D from numerous groups are expected to
lead to a positive result very soon.

We should also note that the losses from coupling to the waveguide
from outside will reduce the number of successful phase shifts per
input number of photons. The losses once inside the waveguide are
much smaller and are of the order of only a few
percent\cite{CCW,MZIs_exp,Poon04,FTTAN05}, i.e. most of the
photons that transverse the device will be phased shifted. Both
loss mechanisms will nevertheless result to a reduction of the
rate of input uncorrelated photons/output of phase shifted
photons. This can be counterbalanced by increasing the rate of
incident photons whenever possible or by integrating a single
photon source in the waveguide. In the case of a complete network
with many gates, some tuning of the individual emitters might be
needed. This is definitely an interesting route to pursue in the
next stage of this project.
 Lastly an implementation of
our scheme requires good synchronization of photon pulses, single
photon sources and very efficient single-photon detectors. These
requirements are very similar to those for quantum computation with
linear optical elements~\cite{KLM}. However, our scheme is much less
demanding in terms of resource overheads per a reliable quantum
gate. Recent progress in the development of single photon sources
indicate that the photonic quantum computation should be a realistic
experimental proposition~\cite{singlephoton}.
 A more detailed study of all possible error mechanisms for this scheme is under way
and will appear elsewhere.

In conclusion, we have discussed how photons propagating in CROWs
could generate strong non-linear interactions which could enable
the implementation of both single and two qubit quantum gates. The
simplicity of the gate switching mechanism using global external
 fields, the feasibility of fabricating two dimensional
photonic crystal structures and CROWs with current or near future
technology and the integrability of this device with optoelectronics
should offer new interesting possibilities for optical quantum
information processing networks\footnote{Shortly after the
submission of this work, we came aware of very recent work where
similar mechanisms for nonlinearities in PBGs are proposed, using
polariton and EIT techniques: I. Friedler et al. quant-ph/0410019 }.

D.G.A. thanks E.N.~Economou, S.N.~Bose and A.~Bychkov for helpful
discussions and D.K.L. Oi for help with the graphs. MFS thanks CNPq
for financial support. This work was supported by the QIP IRC
(GR/S821176/01), and the European Union through the Integrated
 Projects QAP (IST-3-015848), SCALA (CT-015714) and SECOQC. 

\end{document}